\documentclass[aps,prb,twocolumn,superscriptaddress,10pt]{revtex4-1}
\setcitestyle{numbers,square}
\pdfoutput=1

\usepackage[utf8]{inputenc}
\usepackage[T1]{fontenc}
\usepackage[english]{babel} 
\usepackage{amsmath, amsfonts, amssymb}
\usepackage[locale=US,exponent-product = \cdot, separate-uncertainty=true, range-units=single]{siunitx}
\usepackage{xspace}
\usepackage{nicefrac}
\usepackage{multirow}
\usepackage{physics}
\usepackage{graphicx}
\usepackage{enumerate}
\usepackage[dvipsnames]{xcolor}
\usepackage{hyperref}
\hypersetup{
	colorlinks = true,
	linkcolor = blue,
	citecolor = blue,
	filecolor = blue,
	urlcolor = blue
}

\usepackage{prettyref}%
\newcommand{\pref}[1]{\prettyref{#1}}%
\newrefformat{fig}{Fig.~\ref{#1}}%
\newrefformat{tab}{Table~\ref{#1}}%
\newrefformat{sec}{Sec.~\ref{#1}}%
\newrefformat{app}{App.~\ref{#1}}%
\newrefformat{eq}{Eq.~(\ref{#1})}%

\newcommand{\iu}{\ensuremath{\mathrm{i}}} 
\newcommand{\BM}{\ensuremath{{R_1^+}}\xspace}
\newcommand{\eg}{\ensuremath{{e_g}}\xspace}
\newcommand{\ttg}{\ensuremath{{t_{2g}}}\xspace}
\newcommand{\CFO}{CaFeO$_3$\xspace}
\newcommand{\Pton}{\ensuremath{P2_1/n}\xspace}

\begin{document}

\title{Charge disproportionation and Hund's insulating behavior in a five-orbital Hubbard model applicable to $d^4$ perovskites}
\author{Maximilian E. Merkel}
\email{maximilian.merkel@mat.ethz.ch}
\affiliation{Materials Theory, ETH Z\"u{}rich, Wolfgang-Pauli-Strasse 27, 8093 Z\"u{}rich, Switzerland}
\author{Claude Ederer}
\email{claude.ederer@mat.ethz.ch}
\affiliation{Materials Theory, ETH Z\"u{}rich, Wolfgang-Pauli-Strasse 27, 8093 Z\"u{}rich, Switzerland}

\date{\today}

\begin{abstract}
We explore the transition to a charge-disproportionated insulating phase in a five-orbital cubic tight-binding model applicable to transition-metal perovskites with a formal $d^4$ occupation of the transition-metal cation, such as ferrates or manganites. We use dynamical mean-field theory to obtain the phase diagram as a function of the average local Coulomb repulsion $U$ and the Hund's coupling $J$. The main structure of the phase diagram follows from the zero band-width (atomic) limit and represents the competition between high-spin and low-spin homogeneous and an inhomogeneous charge-disproportionated state. This results in two distinct insulating phases: the standard homogeneous Mott insulator and the inhomogeneous charge-disproportionated insulator, recently also termed \emph{Hund's insulator}. We characterize the unconventional nature of this Hund's insulating state.
Our results are consistent with previous studies of two- and three-orbital models applicable to isolated $t_{2g}$ and $e_g$ subshells, respectively, with the added complexity of the low-spin/high-spin transition.
We also test the applicability of an effective two-orbital ($e_g$-only) model with disordered $S=3/2$ $t_{2g}$ core spins. Our results show that the overall features of the phase diagram in the high-spin region are well described by this simplified two-orbital model but also that the spectral features exhibit pronounced differences compared to the full five-orbital description.
\end{abstract}

\maketitle

\section{Introduction}\label{sec:introduction}

Many of the fascinating properties of transition-metal oxides can be understood as a result of the strong Coulomb interaction, experienced by electrons in the rather localized $d$ orbitals, in combination with their mixed ionic-covalent character, which is due to hybridization between the transition-metal $d$ orbitals and the $p$ orbitals of the surrounding oxygen ligands. The resulting electronic structure is then often in the border regime between itinerant and localized, where the electrons experience strong correlations that can give rise to a number of intriguing properties such as, e.g., metal-insulator transitions~\cite{imada_metal-insulator_1998}.

Such correlation effects and the emergence of a Mott-insulating state are usually associated with a large value of the Hubbard parameter $U$, which describes the repulsion between electrons located on the same site in the seminal Hubbard model~\cite{hubbard_electron_1963, kanamori_electron_1963}. However, in multi-band Hubbard models, strong correlations, i.e., pronounced deviations from the behavior of (effectively) non-interacting particles, can also be caused by the Hund's coupling $J$, even if the corresponding system is not particularly close to a Mott-insulating phase~\cite{georges_strong_2013}. Such materials have recently been termed \emph{Hund's metals}~\cite{yin_kinetic_2011, chatzieleftheriou_enhancement_2020, richaud_interaction-resistant_2021}.

Furthermore, a strong Hund's coupling can even lead to the formation of an inhomogeneous, charge-ordered or charge-disproportionated, insulating phase, distinct from the usual (homogeneous) Mott-insulator. The existence of such a charge-disproportionated insulator (CDI) has been demonstrated both for a two-orbital and a three-orbital Hubbard model~\cite{subedi_low-energy_2015, isidori_charge_2019, ryee_nonlocal_2020}, which are applicable to materials with partially filled $e_g$ and $t_{2g}$ sub-shells. Similar physics have also been discussed in the context of valence-skipping metals and insulators resulting from a negative effective Coulomb interaction~\cite{strand_valence-skipping_2014}.

The CDI phase has also been termed \emph{Hund's insulator}~\cite{isidori_charge_2019, ryee_nonlocal_2020, springer_osmates_2020}, to emphasize the crucial role of the Hund's interaction for stabilizing the insulating state and to distinguish it from the usual (homogeneous) Mott-insulator. 
We use the terms CDI and Hund's insulating phase interchangeably throughout this article, but we also note that earlier papers have used the term Hund's insulator in slightly different contexts, e.g., to describe situations with half-filled shells where the Hund's coupling $J$ cooperates with the Hubbard $U$ and the value of $U$ alone is not sufficient to reach an insulating state~\cite{mcnally_hunds_2015} or in specific cases where Hund's coupling leads to the stabilization of a topological insulating phase~\cite{budich_fluctuation-driven_2013}.

The parameter regime where a CDI or Hund's insulator is observed corresponds to an electronic structure where the Coulomb interaction is strongly screened, which leads to a moderate value of the Hubbard parameter $U$, whereas the Hund's parameter $J$ is less affected by the screening. This regime is expected to occur in transition-metal oxides with a pronounced charge transfer character, i.e., strong hybridization between the transition-metal $d$ and oxygen $p$ bands. This can generally be expected to occur towards the end of the 3$d$ transition-metal series or for systems with high oxidation numbers of the transition-metal cation.

The formation of a charge-disproportionated insulating phase has been observed, e.g., in the series of rare earth nickelates, $R$NiO$_3$~\cite{alonso_charge_1999, alonso_room-temperature_2000, mazin_charge_2007, park_site-selective_2012, subedi_low-energy_2015, varignon_complete_2017, peil_mechanism_2019}. Here, the Ni cations are in a formal $d^7$ electron configuration, which then disproportionates according to $2d^7 \rightarrow d^6 + d^8$ (or, equivalently, $2d^8\underline L \rightarrow d^8\underline L^2 + d^8$, where $\underline L$ denotes a ligand hole in the surrounding oxygen network~\cite{mizokawa_spin_2000, johnston_charge_2014}). 
In this case the $t_{2g}$ subshell is always completely filled with six electrons, and the remaining Ni-$d$ electron occupies the $e_g$ subshell.
It has been shown that the underlying physics can be well described within a minimal two-orbital $e_g$ Hubbard model with an average quarter-filling~\cite{subedi_low-energy_2015}. 
Furthermore, the charge disproportionation lowers the symmetry of the crystal structure and couples strongly with a lattice distortion corresponding to a \emph{breathing mode} of the octahedral network with $R_1^+$ symmetry, resulting in a three-dimensional checkerboard-like pattern of alternating small and large oxygen octahedra surrounding the nominal $d^6$ and $d^8$ Ni sites, respectively.

Similar behavior has also been observed for the alkaline earth ferrate \CFO~\cite{takano_charge_1977, takeda_preparation_1978, kawasaki_phase_1998, woodward_structural_2000, takeda_metalsemiconductor_2000} and in PbCrO$_3$~\cite{wu_pressure-induced_2014, cheng_charge_2015}. The case of PbCrO$_3$ is potentially more complex due to the presence of the stereochemically active Pb$^{2+}$ cation. However, the nominal $d^2$ configuration of the Cr$^{4+}$ cation and the suggested charge disproportionation according to $3\rm{Cr}^{4+} \rightarrow 2\rm{Cr}^{3+} + \rm{Cr}^{6+}$ is consistent with recent theoretical predictions for the three-orbital Hubbard model~\cite{isidori_charge_2019}.
On the other hand, the Fe$^{4+}$ cation in \CFO is assumed to be in a nominal $d^4$ high spin configuration~\cite{takano_pressure-induced_1991}, 
with partial filling of both $t_{2g}$ and $e_g$ subshells, and thus requires a description using the full five-orbital $d$ manifold. Additionally, the competition between high-spin and low-spin states might introduce further complexity in the underlying physics. 

Motivated by this, we study a full five-orbital Hubbard model with $d^4$ electron filling. We note that, while our main motivation is to study the possible emergence of a charge disproportionated state in a rather generic five-orbital TB model, we fit the parameters of this model to the actual bandstructure of CaFeO$_3$ in order to work with realistic values for bandwidth and crystal-field splittings. Thus, we consider the physically relevant case with cubic symmetry by incorporating a crystal-field splitting between the three-fold degenerate lower-lying $t_{2g}$ states and the two-fold degenerate higher-lying $e_g$ states.
We use dynamical mean-field theory (DMFT) to calculate expectation values of this model around room temperature. This allows us to identify all relevant phases as function of the local interaction parameters $U$ and $J$, describing the average Coulomb repulsion and Hund's coupling within the $d$-shell. 

In particular, we establish the presence of a spontaneously charge-disproportionated insulating phase in this five-orbital model, in addition to the usual homogeneous Mott-phase. The overall structure of the phase diagram resembles analogous results for the two- and three-orbital models, but with the additional feature of low-/high-spin transitions.
We analyze the character of the insulating state, and we explore the effect of a structurally induced energy difference between the $e_g$ states on the two different sub-lattices as well as the effect of a reduced $e_g$ band width.
Furthermore, we compare the phase diagram obtained for the full five-orbital model, with an effective two-orbital model, applicable to the high-spin limit~\cite{ahn_effects_2000, pavarini_origin_2010}. This simplified model is able to quantitatively reproduce many features of the full five-orbital model, including, e.g,  the phase boundaries as function of the interaction parameters, but also exhibits pronounced differences in the spectral properties. 

\section{Model and methods}

In this section, we first introduce the models that we are studying: the complete five-orbital model for the full $d$ shell and the effective two-orbital $e_g$ model with localized $t_{2g}$ spins. 
We then describe how the parameters of the models are fitted to the band structure of CaFeO$_3$ obtained from density functional theory (DFT). Finally, we describe the details of our DMFT calculations and introduce the observables used to characterize the different phases.

\subsection{The Hamiltonian}
\label{sec:ham}

We study the following model, which consists of a single-particle tight-binding (TB) part and a local interaction on each site $R$:
\begin{equation}
\label{eq:ham}
    H  = H_\text{TB} + \sum_R H_\text{int}^{(R)} \quad .
\end{equation}
The TB part consists of inter-site hopping terms as well as an on-site crystal-field splitting, $\Delta_{\eg - \ttg}$, between \eg and \ttg orbitals. We include hopping between nearest neighbors (NN) and next-nearest neighbors (NNN) on a cubic lattice, representing the sites of the transition-metal cations in the perovskite structure. The hopping amplitudes are assumed to have the cubic symmetry corresponding to \ttg and \eg orbitals. 
Thus, between nearest neighbors there is only hopping between \ttg orbitals of the same type and no hopping connecting the \eg and \ttg orbitals. 
For simplicity, we neglect next-nearest neighbor hopping between different $t_{2g}$ and between $t_{2g}$ and $e_g$ orbitals, and we assume the same fixed ratio $t_\mathrm{NNN}/t_\mathrm{NN}$ between the NNN and NN hopping amplitudes for both \eg and \ttg orbitals. 
For the $e_g$ orbitals, $t_\text{NN}$ and $t_\text{NNN}$ correspond to the hopping between $d_{3z^2-r^2}$ orbitals along $z$ and along $x+z$ or $y+z$, respectively, see Ref.~\onlinecite{ederer_structural_2007}.

We also perform calculations where we introduce an intrinsic ``site splitting'' of the $e_g$ levels, by raising, respectively lowering, the local on-site energy of the \eg orbitals on adjacent sites by $\pm\Delta_\eg/2$.
This mimics the effect of the structural $R_1^+$ breathing mode distortion that typically accompanies the charge disproportionation in perovskites (see, e.g., Ref.~\onlinecite{hampel_interplay_2019}). As described in \pref{sec:introduction}, this distortion consists in a spatially alternating expansion and compression of the oxygen octahedra surrounding the transition-metal sites, resulting in a three-dimensional checkerboard-like pattern of large and small octahedra. 

To describe the local electron-electron interaction within the five-orbital $d$ shell, we use the so-called Slater parametrization in the density-density approximation (see, e.g., Ref.~\onlinecite{pavarini_ldadmft_2011}):
\begin{align}
\label{eq:hint}
    H_\mathrm{int} = &\frac12 \sum_{mm',\sigma} U_{mm'} n_{m\sigma} n_{m'\bar\sigma} \nonumber\\
    + &\frac12 \sum_{m\neq m',\sigma} (U_{mm'} - J_{mm'}) n_{m\sigma} n_{m'\sigma} \quad .
\end{align}
This term is purely local and identical for each transition-metal site (we therefore suppress the site index $R$ in~\pref{eq:hint}).
Thereby, $n_{m\sigma}$ is the occupation number operator for orbital $m$ with spin $\sigma$ (on site $R$) and $\bar{\sigma}$ is short for $-\sigma$.
The matrix elements $U_{mm'}$ and $J_{mm'}$ are parametrized in terms of the Slater integrals $F_0$, $F_2$, and $F_4$ in the usual form~\cite{pavarini_ldadmft_2011}. The Slater integrals themselves are expressed in terms of the average Coulomb repulsion $U = F_0$ and Hund's rule interaction $J = (F_2 + F_4)/14$, using a fixed ratio $F_4/F_2 = 0.63$~\cite{pavarini_ldadmft_2011}. 

\subsection{The effective two-orbital approximation}
\label{sec:two-orbital}

We also compare the full five-orbital $d$-shell model to a simplified effective two-orbital \eg model, applicable to the high-spin limit. In this model, the occupation of the \ttg states on each site is fixed to three electrons, giving rise to an $S=3/2$ \emph{core spin}, and hopping in and out of these \ttg states is neglected~\cite{ahn_effects_2000,pavarini_origin_2010}. The remaining electron is thus constrained to the $e_g$ orbitals, but can still hop from site to site.
Then, for every site, the \ttg core spin defines a local spin quantization axis, and the spins of the \eg  electrons are defined relative to this local axis, denoted by $\sigma=\Uparrow$/$\Downarrow$. 

The $e_g $ electrons interact with the \ttg core spins through an effective Hund's rule interaction, which leads to a local spin splitting of the \eg orbitals:
\begin{align}
    H = -h \sum_m (n_{m\Uparrow} - n_{m\Downarrow}),
\end{align}
where $h$ is the effective magnetic field created by the \ttg core spin. The strength of this effective magnetic field can be related to the parameters of the full interaction Hamiltonian, \pref{eq:hint}, in particular to the Hund's interaction parameter $J$, by considering the energy gain (penalty) of an \eg spin (anti-)parallel to the \ttg core spin, i.e., by considering the energy difference between $(t_{2g})^3_\uparrow (e_g)^1_\uparrow$ and $(t_{2g})^3_\uparrow (e_g)^1_\downarrow$:
\begin{align}
    2h = \frac6{49} F_2 + \frac{25}{147} F_4 \approx 1.97 J \quad .
\end{align}

Furthermore, assuming a completely random, i.e., paramagnetic orientation of the \ttg core spins on the different sites results in a renormalization of the hopping amplitudes, enabling also hopping between $\Uparrow$ and $\Downarrow$ states corresponding to different sites.
After averaging over all possible relative spin orientations, this can be described by a matrix $u^{R, R'}_{\sigma, \sigma'}$, which does not affect the on-site Hamiltonian ($u^{R, R}_{\sigma, \sigma'} = \delta_{\sigma, \sigma'}$) but renormalizes and mixes the inter-site hopping terms ($u^{R, R'\neq R}_{\sigma, \sigma'} = 2/3$) according to~\cite{ahn_effects_2000,pavarini_origin_2010}:
\begin{align}
    t^{R, R'}_{m, m'} \rightarrow t^{R, R'}_{m, m'} u^{R, R'}_{\sigma, \sigma'} \quad .
\end{align}

To enable a consistent comparison with the full five-orbital model, the interaction in the effective two-orbital model is described using the \eg subspace of the full interaction Hamiltonian,  \pref{eq:hint}. This is equivalent to the density-density approximation of the so-called Kanamori Hamiltonian~\cite{kanamori_electron_1963}, but with the corresponding interaction parameters defined in terms of the Slater integrals (or in terms of the average Coulomb repulsion $U$ and the corresponding Hund's rule interaction $J$). 

\subsection{Fitting the parameters of the TB model}
\label{sec:fitting}

The purpose of this work is to study the physics of charge disproportionation in a relatively generic cubic five-orbital model. Nevertheless, we focus on realistic parameter regimes, applicable to perovskite transition-metal oxides and closely related materials. We therefore obtain parameters for the single-particle part of the Hamiltonian, $H_\text{TB}$, by comparison with the full band structure of CaFeO$_3$, calculated within density functional theory (DFT).

We perform DFT calculations using the ``Vienna Ab-initio Simulation Package'' (VASP)~\cite{kresse_ab_1993, kresse_efficient_1996} employing the PBEsol exchange-correlation functional~\cite{perdew_restoring_2008}, both for the high-temperature $Pbnm$ structure, where all Fe sites are equivalent, and the charge-disproportionated \Pton crystal structure observed below \SI{290}{K}~\cite{morimoto_structure_1996, takeda_metalsemiconductor_2000}. 
We use a $7 \times 7 \times 5$ $k$-point mesh for the 20-atom unit cell and a plane-wave energy cutoff of \SI{600}{eV} to ensure well converged results.

We first relax unit-cell parameters as well as atomic positions within $Pbnm$ symmetry. In order to stabilize the high-spin state of the Fe cation, a Hubbard-corrected DFT+$U$ treatment and the presence of antiferromagnetic (AFM) order is required.
We find that moderate values of $U=\SI{4}{eV}$ and $J=\SI{1}{eV}$ and A-type antiferromagnetic order (as proxy for the more complicated helical spin structure in \CFO~\cite{rogge_itinerancy-dependent_2019}) result in structural parameters that are in good agreement with experimental data (e.g., the calculated unit cell volume is \SI{210.34}{\angstrom\cubed})~\cite{morimoto_structure_1996, woodward_structural_2000}.  
We then re-relax the atomic positions within non-spin-polarized DFT (to exclude any structural effects or symmetry lowering stemming from the magnetic order) but with lattice parameters fixed to the ones obtained from the magnetic $+U$ calculations. 
We note that a non-magnetic DFT calculation leads to a low-spin state of the Fe cation, which would result in an unrealistically low unit-cell volume.

To obtain a realistic estimate of the site-splitting $\Delta_{e_g}$, we manually add a breathing mode of $\BM = \SI{.2}{\angstrom}$, very similar to the experimentally observed amplitude $\BM = \SI{.18}{\angstrom}$~\cite{woodward_structural_2000}, while keeping all other structural parameters unchanged.
A full DFT+$U$ study of CaFeO$_3$ has been presented in Ref.~\onlinecite{zhang_charge-_2018}. 

\begin{figure}
    \centering
    \includegraphics[width=1\linewidth]{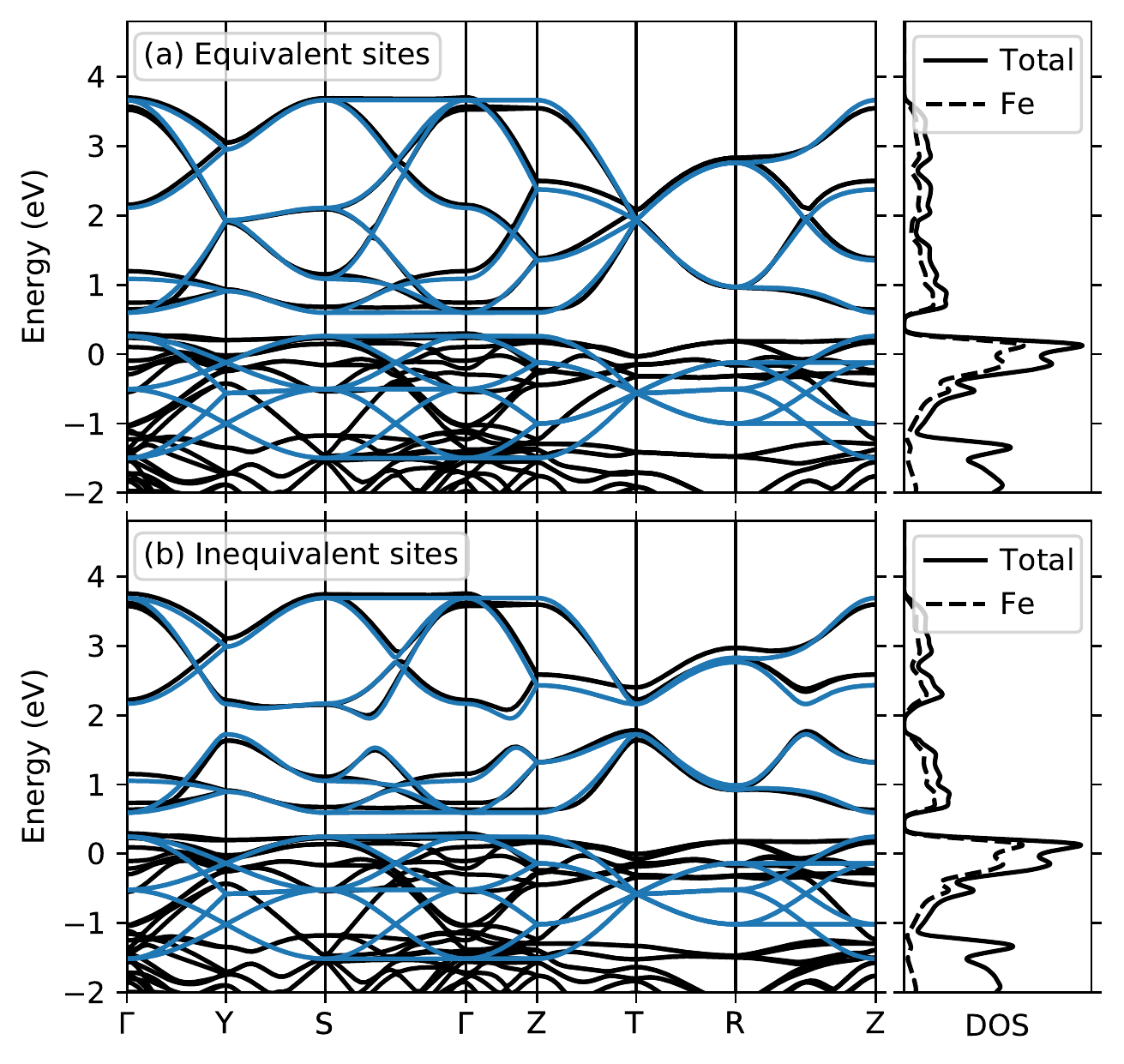}
    \caption{Plots of the band structure of \CFO obtained using DFT (black lines) and the fitted TB bands (blue lines), in (a) $Pbnm$ symmetry ($\BM = \SI{.0}{\angstrom}$) and (b) \Pton symmetry ($\BM = \SI{.2}{\angstrom}$). Zero energy indicates the Fermi level.
    Additionally, the DFT total density of states (DOS) and Fe-projected DOS are shown in the corresponding right panels.}
    \label{fig:bands_dft_tb}
\end{figure}

The calculated band structure and corresponding densities of states (DOS) in the relevant energy window around the Fermi level are shown in \pref{fig:bands_dft_tb}. It can be seen that a group of bands with dominant Fe character, located between 0.5\,eV and 4\,eV above the Fermi level, is separated from all other bands at higher and lower energies. Further inspection confirms that these are the Fe-\eg bands. The bands between approximately $-$1\,eV and 0.5\,eV also have mostly Fe character and correspond to the Fe-\ttg states. Towards lower energies, there is some entanglement of these Fe-\ttg bands with other bands with dominant O-$p$ character.
Comparing the band structures obtained with and without the structural $R_1^+$ breathing mode, one can see that its main effect is to introduce a splitting throughout the middle of the \eg bands, analogously to what has been observed for nickelates~\cite{subedi_low-energy_2015, mercy_structurally_2017}. In contrast, the Fe-\ttg bands are not much affected by the structural distortion.

We fit the parameters of our TB model such that the corresponding band dispersion matches the bands obtained from our DFT calculations. Thereby, the ratio between NNN and NN hopping, $t_\text{NNN}/t_\text{NN}$, is obtained by fitting the well-separated \eg bands, whereas the \ttg NN hopping is then mainly adjusted to obtain the correct \ttg band width. Note that the two cases with and without breathing mode are fitted with the same hopping amplitudes, whereas the average \eg-\ttg splitting $\Delta_{\eg - \ttg}$ is slightly altered by the \BM{} breathing mode. 
The resulting TB band structure is also shown in \pref{fig:bands_dft_tb}, and the corresponding parameters are listed in \pref{tab:tb_parameters}. It can be seen that the TB model results in an excellent fit of the Fe-\eg bands, both in the $Pbnm$ and $P2_1/n$ cases. For the \ttg bands, the overall band width is well described, while the dispersion of the individual bands shows notable deviations. This could be improved by introducing a separate $t_\text{NNN}/t_\text{NN}$ for the \ttg hopping or other additional intra- and inter-site TB parameters. However, for simplicity, and since the specific form of the \ttg bands is not relevant in the context of this work, we refrain from introducing further parameters into our TB model.

\begin{table}
    \centering
    \caption{The parameters of our TB model, obtained by fitting to the DFT band structures of \CFO with and without the \BM breathing mode, together with the corresponding $e_g$ and $t_{2g}$ total band widths, $W_{e_g}$ and $W_{t_{2g}}$, respectively.}
    \label{tab:tb_parameters}
    \begin{ruledtabular}
    \begin{tabular}{lcc}
        & $\BM{}=0.0$\,\AA & $\BM =0.2$\,\AA \\
        \hline
        $\Delta_{\eg - \ttg}$ (\si{eV}) & 2.494 & 2.519 \\
        $t_\eg$ (\si{eV}) & $-$0.511 & $-$0.511 \\
        $t_\ttg$ (\si{eV}) & $-$0.220 & $-$0.220 \\
        $t_\text{NNN}/t_\text{NN}$ & 0.065 & 0.065 \\
        $\Delta_\eg$ (\si{eV}) & 0 & 0.436 \\
        \hline
        $W_\eg$ (\si{eV}) & \multicolumn{2}{c}{3.069} \\
        $W_\ttg$ (\si{eV}) & \multicolumn{2}{c}{1.761} 
    \end{tabular}
\end{ruledtabular}
\end{table}

\subsection{DMFT calculations}
\label{sec:dmft}

We obtain the phase diagram of the model, \pref{eq:ham}, as function of the interaction parameters $U$ and $J$ using dynamical mean-field theory (DMFT)~\cite{georges_dynamical_1996}. Within DMFT, the self-energy of the full lattice problem is approximated by a frequency-dependent but local self-energy, which is obtained by mapping each symmetry-inequivalent site on an effective impurity problem, defined via a self-consistency condition requiring that the impurity Green's function is identical to the corresponding local Green's function of the full lattice problem.

The calculations are implemented using the TRIQS and DFTTools libraries~\cite{parcollet_triqs_2015, aichhorn_triqsdfttools_2016, solidmft}.
To solve the effective impurity problem, we use a continuous-time Quantum-Monte-Carlo solver (CT-HYB), also implemented in the TRIQS library~\cite{seth_triqscthyb_2016}, and an inverse temperature $\beta=1/(k_\text{B}T)=\SI{40}{eV^{-1}}$, corresponding to approximately room temperature. Green's functions are represented by 40 Legendre coefficients~\cite{boehnke_orthogonal_2011} for the full five-orbital model (50 for the effective two-orbital model), which results in good convergence and a smooth behavior of the self-energy in imaginary frequency $\Sigma(\iu\omega_n)$. 

We average over both spin channels and also over the \eg and \ttg manifolds separately
to ensure a paramagnetic solution without orbital polarization in the two different subshells.
Furthermore, to allow for a charge disproportionated solution, we divide the sites of the simple cubic lattice into two interpenetrating fcc lattices, (I) and (II), corresponding to the expected 3D checkerboard-like arrangement, and solve the two resulting effective impurity problems separately. 

From the local Green's function \mbox{$G_{mm'}(\tau) = - \langle T c_m(\tau) c^\dagger_{m'}(0) \rangle$}, where $T$ is the imaginary-time ordering operator and $c^\dagger_m(\tau)$ is the creation operator for an electron in orbital $m$ at the local site at imaginary time $\tau$, 
we obtain the total \eg occupation per site, $n_\mathrm{e_g} = n_\mathrm{z^2} + n_\mathrm{x^2-y^2}$, the (positively-defined) occupation difference between the two sub-lattices, $\delta n = |\sum_m n_m^\mathrm{(I)} - \sum_m n_m^\mathrm{(II)}|$, 
and the averaged spectral weight around the Fermi level, \mbox{$A(\omega=0) = - (\beta/\pi) \mathrm{Tr}\, G(\tau=\beta/2)$}.
Furthermore, we obtain an estimate for the (orbital-dependent) quasiparticle weight $Z_m = \left[1 - \left. \frac{\partial \Im \Sigma_m(\iu\omega)}{\partial (\iu\omega)} \right|_{\iu\omega=0}  \right]^{-1}$ by fitting a fourth-order polynomial to the imaginary part of the local self-energy, $\Im \Sigma_m (\iu\omega_n)$, for the eight lowest positive Matsubara frequencies $\iu\omega_n$ and extrapolate the slope at zero frequency from this fit, {\it c.f.} Appendix C in Ref.~\onlinecite{schafer_tracking_2021}.
We use the maximum-entropy method~\cite{jarrell_bayesian_1996}, implemented in the TRIQS library~\cite{maxent}, to obtain both $k$-averaged and $k$-resolved spectral functions on the real frequency axis.

\section{Results}\label{sec:results}

\subsection{Atomic-limit predictions}
\label{sec:atomic-limit}

Before presenting our results from the full DMFT calculations, we outline the general features of the expected phase diagram by considering the ``atomic limit'' of \pref{eq:ham}, i.e., the limit where all hopping amplitudes and thus the associated band widths go to zero, and the ground state configuration of the system is solely determined by the local part of the Hamiltonian, i.e., the interaction term $H_\text{int}$ plus the local crystal-field and \eg site splitting.

\begin{figure}
    \centering
    \includegraphics[width=1\linewidth]{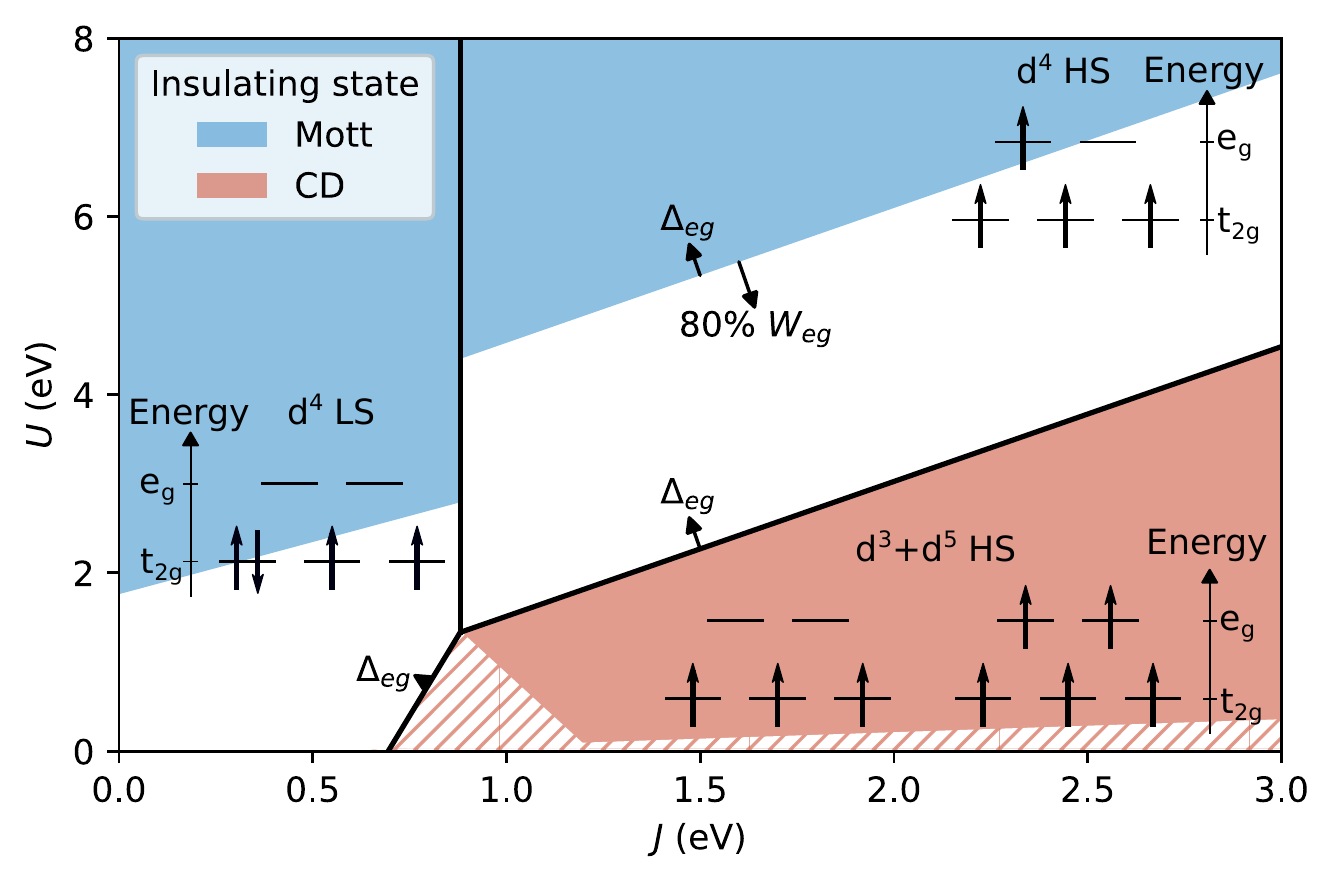}
    \caption{Schematic phase diagram for our TB model corresponding to equivalent sub-lattices, $\Delta_{e_g}=0$. Thick black lines separate regions with different electronic ground states in the atomic limit: homogeneous HS $d^4$, homogeneous LS $d^4$, and inhomogeneous HS $d^3$ + $d^5$, as indicated by the energy level diagrams. The colored regions indicate the expected insulating regions, with the homogeneous Mott phase in blue and the charge disproportionated (CD) insulating region in red. The white area represents the expected homogeneous metallic state. The red-hatched area denotes a potential CD metallic region based on simple band-width estimates (see main text). 
    The shifts of the phase boundaries resulting from a site splitting $\Delta_{e_g}=\SI{.436}{eV}$ and an 80\,\% reduction of the $e_g$ band with $W_{e_g}$ are indicated by small black arrows, with the length of the arrow corresponding to the actual shift.}
\label{fig:schematic_phase_diagram}
\end{figure}

We first compare the energies of the three most relevant configurations: (i) homogeneous high-spin (HS) $d^4$; (ii) homogeneous low-spin (LS) $d^4$; and (iii) inhomogeneous HS $d^5$ + $d^3$, also depicted by the schematic level diagrams in \pref{fig:schematic_phase_diagram}, and identify the ground state as function of interaction parameters $U$ and $J$. Details are described in \pref{app:atomic_limit}.
\pref{fig:schematic_phase_diagram} depicts the case without an explicit \eg site splitting, i.e., $\Delta_\eg=0$. Thick black lines separate the regions where the three different cases (i)-(iii) are the ground state configuration.
The homogeneous LS configuration (ii) is generally favorable for small $J$ ($J \lesssim \Delta_{e_g-t_{2g}}/2.83 \approx 0.88$\,eV, except for the small part separating the LS and inhomogeneous HS regions, where the critical $J$ is further decreased by decreasing $U$). The inhomogeneous charge-disproportionated (CD) state (iii) is the ground state for larger $J$ and $U\lesssim 1.51J$ (except, again, for the small part separating CD and homogeneous LS regions, where the critical $U$ is smaller). For large $J$ and $U$, the homogeneous HS state (i) is obtained.
We note that in the Slater parametrization of the local interaction, a positive $U$ represents an \emph{average repulsive} electron-electron interaction (independent of $J$), which is why we consider the whole region with $U>0$ as potentially physically relevant.

A simple estimate for the boundary between metallic and (Mott-) insulating regions can be obtained by comparing the lowest inter-site charge transfer excitation energy with the average electronic kinetic energy. For the homogeneous phases, this kinetic energy can be approximated by the relevant band width (see, e.g., Ref.~\onlinecite{georges_strong_2013}), resulting in the blue-shaded ``Mott-insulating'' region in \pref{fig:schematic_phase_diagram} (see \pref{app:atomic_limit} for details).
Since the \ttg band width is smaller than the \eg band width, the Mott phase extends to smaller $U$ values in the LS region than in the HS region.

For the CD region, a good estimate for the corresponding effective band width/kinetic energy is more difficult to obtain. For the transition to the homogeneous metallic phase it turns out to be significantly smaller than $W_\eg$ and essentially goes to zero when $\Delta_\eg$ becomes large~\cite{subedi_low-energy_2015}. For simplicity, we therefore denote the whole region below the CD transition line as ``CDI'' in the atomic limit (red shaded area in \pref{fig:schematic_phase_diagram}). Considering only hopping among either nominal $d^5$ or $d^3$ sites defines limits for the CDI region towards small $U$ (see \pref{app:atomic_limit}). Simple band-width estimates for the kinetic energy result in the limits indicated by the red-striped regions in \pref{fig:schematic_phase_diagram}.

Introducing a nonzero \eg site splitting  $\Delta_\eg$
(corresponding to a nonzero \BM distortion of the underlying crystal structure), extends the CD region slightly by rigidly shifting the phase boundaries towards the homogeneous HS and LS phases (see \pref{app:atomic_limit}) as indicated by the small arrows in \pref{fig:schematic_phase_diagram}. This also shifts the metal/insulator phase boundary within the homogeneous HS region towards higher $U$.
In contrast, reducing the \eg band width shifts the metal/insulator phase boundary in the homogeneous HS phase in the opposite direction, while not affecting any of the other main phase boundaries. 

\subsection{Phase diagram for equivalent sites}

\begin{figure}
    \centering
    \includegraphics[width=1\linewidth]{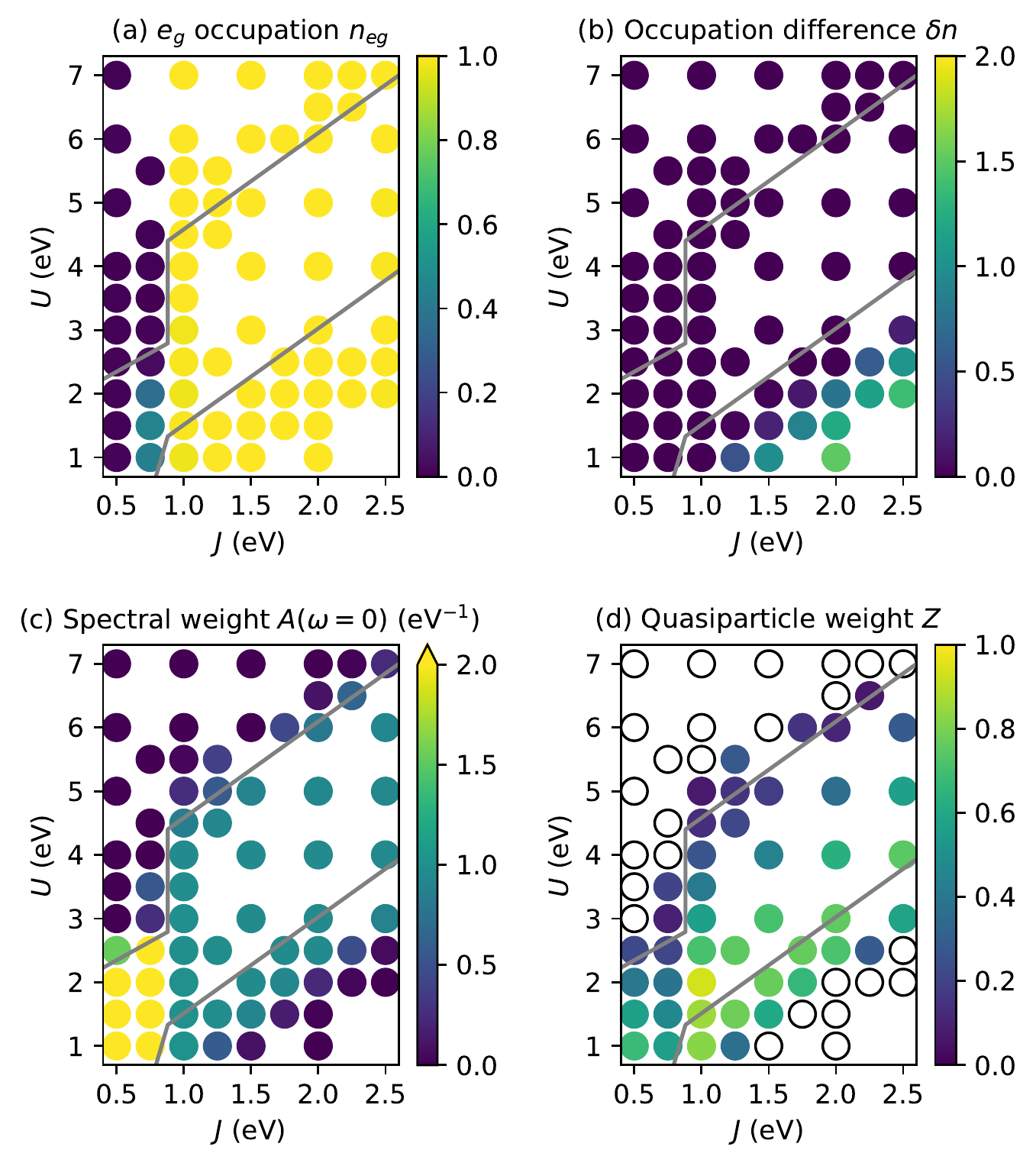}
    \caption{The characteristic properties for the phase diagram in the basic TB model with equivalent sites: (a) average $e_g$ occupation $n_{e_g}$, (b) occupation difference $\delta n$ between the two sub-lattices, (c) averaged spectral weight $A(0)$, and (d) quasiparticle weight $Z$, estimated from the slope of $\Im\Sigma(\iu\omega)$. Atomic limit boundaries are indicated as solid gray lines. In (d), the empty circles indicate data points where the system is insulating and thus $Z$ is ill-defined.}
    \label{fig:characteristics_noBM_fullBW}
\end{figure}

Next, we present the results of our full DMFT calculations for the TB model with $\Delta_{e_g} = 0$, i.e., for the case without a site-dependent shift of the $e_g$ levels. \pref{fig:characteristics_noBM_fullBW} depicts various observables as function of interaction parameters $U$ and $J$. 

The total $e_g$ occupation per site, $n_{e_g}$, shown in \pref{fig:characteristics_noBM_fullBW}(a), indicates a sharp transition from LS ($n_{e_g}=0$) to HS ($n_{e_g}=1$) for $J \geq 1$\,eV, essentially independent of $U$, in very good agreement with the expectation obtained from the atomic limit discussed in \pref{sec:atomic-limit}. Only for $J=0.75$\,eV and $U \leq 2$\,eV, we obtain a small intermediate-spin region where the $e_g$ occupation lies between 0 and 1.

The most important result is visible in the occupation difference between the two sublattices, $\delta n$, shown in \pref{fig:characteristics_noBM_fullBW}(b).
Here, in the region of small $U$ but large $J$, i.e., in the region where the atomic limit predicts the inhomogeneous CD ground state, we indeed obtain a non-zero value for $\delta n$, indicating that the system exhibits a spontaneously symmetry-broken state with higher and lower $e_g$ occupation on the two sublattices. Note that this CD region lies fully within the HS region where the average $e_g$ occupation is equal to 1.
Inspection of the spectral weight at zero energy, $A(0)$, 
shows that the charge disproportionation coincides with a transition to an insulating state, indicated by $A(0) \approx 0$ (dark regions in \pref{fig:characteristics_noBM_fullBW}(c)). 
This CD insulating region is separated from the Mott-insulating regime, where $\delta n=0$ and which occurs for large $U$, by a metallic region. It can be seen that the boundary between the Mott-insulating and the metallic region (both in the HS and LS regime) also matches very well with the atomic limit prediction from \pref{sec:atomic-limit}, indicated by the gray lines in \pref{fig:characteristics_noBM_fullBW}.
On the CD side, the MIT occurs together with the emerging charge disproportionation for $\delta n \gtrsim 1$, somewhat below the homogeneous-inhomogeneous transition line predicted from the atomic limit. However, we note that the distance of the atomic-limit prediction and the actual charge disproportionation is significantly less than the $e_g$ band width.

\begin{figure}
    \centering
    \includegraphics[width=1\linewidth]{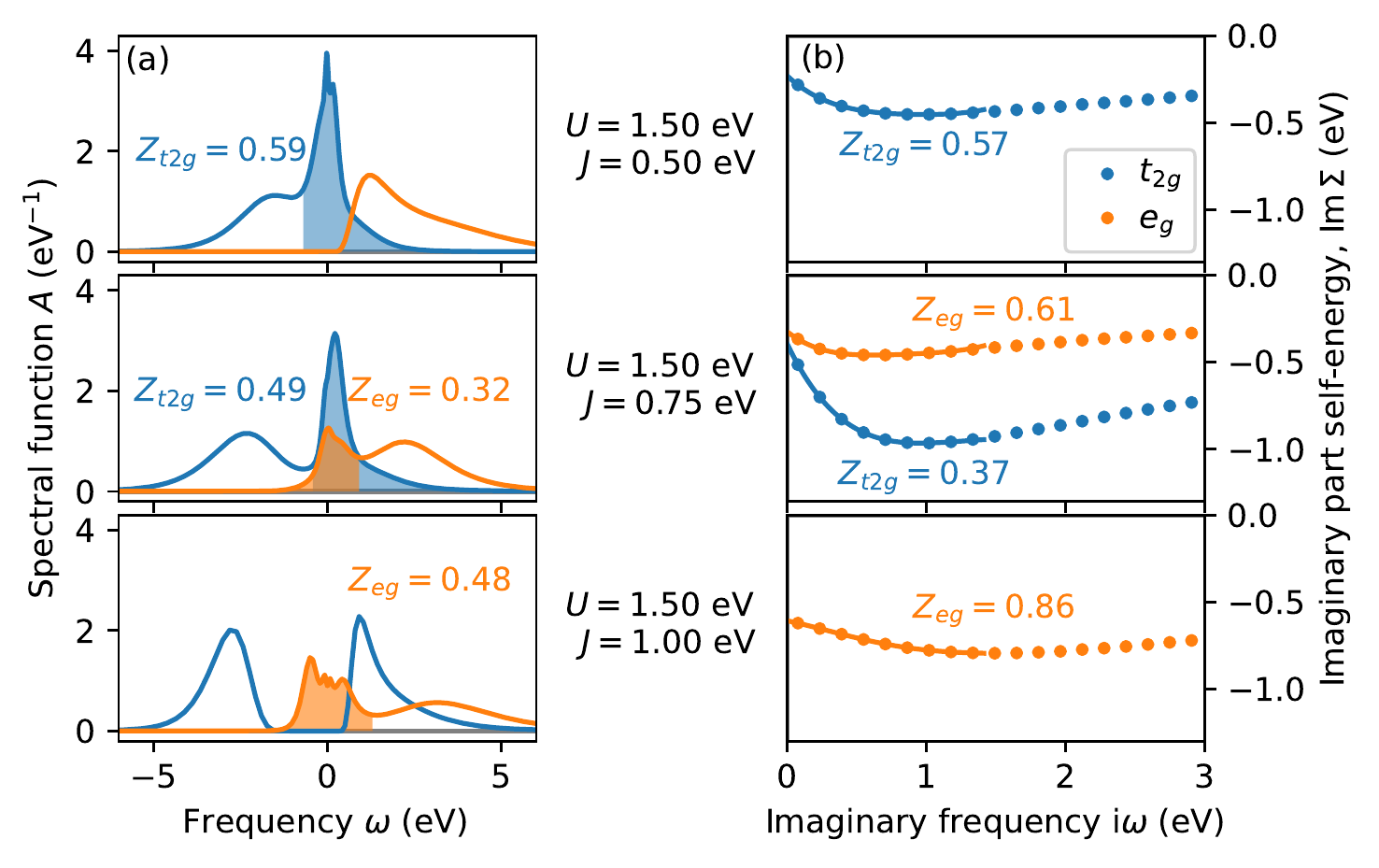}
    \caption{Evolution of the $e_g$- and $t_{2g}$-resolved spectral functions (a) and corresponding imaginary parts of the self energy on the Matsubara axis (b), for the transition from low- (top, $J=0.5$\,eV) to intermediate- (middle, $J=0.75$\,eV) to high-spin (bottom, $J=1.0$\,eV) in the metallic region of the phase diagram (for $U=1.5$\,eV). Also indicated are quasiparticle weights $Z$ estimated by integrating $A(\omega)$ over the shaded areas in (a) or obtained from the slope of $\Im\Sigma(\iu\omega\rightarrow 0)$ (with the corresponding fourth order fits shown as solid lines) in (b).}
    \label{fig:quasiparticle_weight}
\end{figure}

In \pref{fig:characteristics_noBM_fullBW}(c), there is a significant change in spectral weight, $A(0)$, at the LS/HS transition in the metallic region. \pref{fig:quasiparticle_weight}(a) shows corresponding spectral functions (for $U=1.5$\,eV), separated in $t_{2g}$ and $e_g$ contributions. It can be seen that for $J=0.5$\,eV the full spectral weight is generated by the $t_{2g}$ contribution, while the $e_g$ states are still unoccupied. Around $J=0.75$\,eV the transition occurs, where both $e_g$ and $t_{2g}$ states contribute to the spectral weight at zero energy. Above this LS/HS transition, at $J=1$\,eV, the $t_{2g}$ spectral function is gapped, and $A(0)$ is entirely due to the $e_g$ states, i.e., a situation that corresponds to an orbitally selective Mott state. Here, the $e_g$ quasiparticle peak is lower and broader compared to the $t_{2g}$ peak for $J=0.5$\,eV, which results in the lower spectral weight in the phase diagram for the HS case.

\pref{fig:characteristics_noBM_fullBW}(d) shows the quasiparticle weight $Z$ in the metallic region, averaged over all orbitals with non-vanishing spectral weight. $Z$ is largest along the line separating homogeneous/inhomogeneous ground states in the atomic limit, and decreases towards both insulating phases. This is consistent with analogous results for the three-orbital case in Ref.~\onlinecite{isidori_charge_2019}, where this behavior was related to the competition between the two different insulating phases, resulting in the so-called ``Janus effect''~\cite{de_medici_janus-faced_2011, georges_strong_2013}.

\pref{fig:quasiparticle_weight}(b) depicts the imaginary part of the self-energy on the Matsubara axis across the LS/HS transition within the metallic regime. As described in \pref{sec:dmft}, the quasiparticle weight $Z$ is obtained by fitting the low-frequency behavior of $\Im\Sigma(\iu\omega)$. The corresponding fits are shown in \pref{fig:quasiparticle_weight}(b). 
It can be seen that the self-energies exhibit rather strong deviations from simple Fermi liquid behavior. A linear dependence on $\iu\omega$ can only be observed in a very small range close to $\omega \rightarrow 0$, and there is a rather large non-vanishing imaginary part for zero frequency. This could in part be due to the relatively high temperature ($\beta=40\,$eV, corresponding to approximately room temperature) in our calculations.
Thus, the interpretation of $[1-\partial \Im \Sigma/\partial(\iu\omega)]^{-1}$ as effective quasiparticle renormalization is somewhat approximate, here. This can also be inferred from the comparison with the ``quasiparticle weights'' extracted by integrating over the central peak of the corresponding spectral functions in \pref{fig:characteristics_noBM_fullBW}. (Note that the latter of course also just represents a very rough and approximate measure of the quasiparticle renormalization.) 

\begin{figure}
    \centering
    \includegraphics[width=1\linewidth]{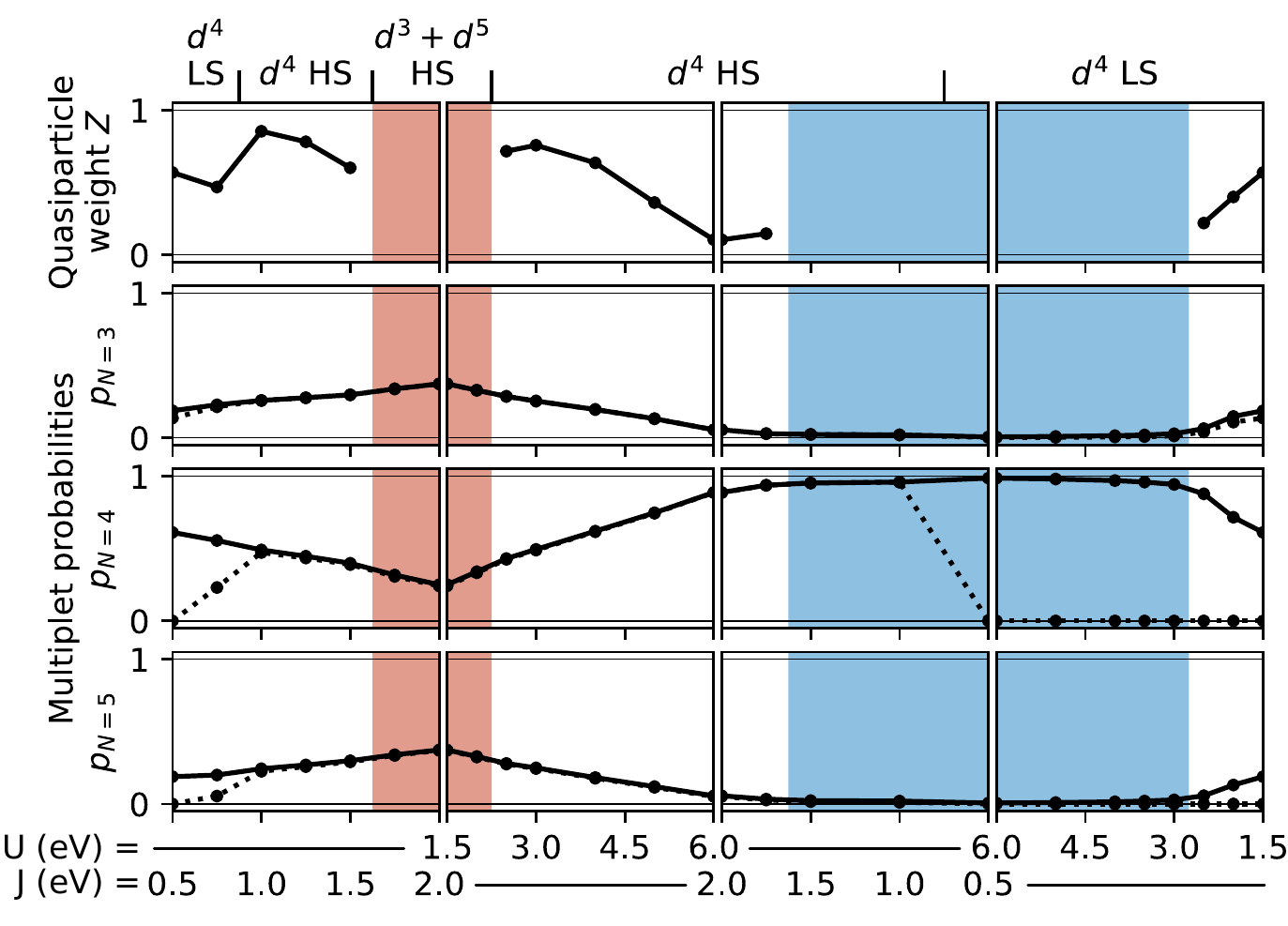}
    \caption{Multiplet analysis for the model with equivalent sites, sampled along a rectangular path in the phase diagram, as described in the main text. For the different segments of this path, either $U$ or $J$ are kept constant while the other parameter is varied as indicated on the $x$-axis. For the multiplet probabilities $p_{N=3}$, $p_{N=4}$, $p_{N=5}$, the solid lines correspond to the sum over all spin quantum numbers $|m_s|$ and the dotted lines to the sum over only the HS multiplets, i.e., those with maximal spins $|m_s| = N/2$. All quantities are averaged over the two sublattices. Blue, red, and white background represents Mott-insulating, CDI, and metallic regions expected from the atomic limit.}
    \label{fig:multiplets_noBM_fullBW}
\end{figure}

To obtain further insight into the different phases, we also analyze the probabilities of the different atomic multiplet states, obtained from the DMFT Quantum-Monte-Carlo solver. These probabilities describe how likely it is for the corresponding state to be visited during the simulation of the effective impurity problem and allow to distinguish between different occupations and LS/HS configurations in a more quantitative way than in the phase diagrams shown in \pref{fig:characteristics_noBM_fullBW}.
To simplify the analysis, we sum up the probabilities for Hilbert states with the same number of electrons $N$ and the same magnetic spin quantum number $|m_s|$~\footnote{We note that within the density-density approximation for the interaction Hamiltonian, \pref{eq:hint}, the magnetic quantum number $m_s$ and not the total spin $S$ is the relevant quantum number indicating the spin state of the system.}.
It turns out that only states with $N=3$, $4$, and $5$ have significant probabilities; these together with the quasiparticle weight $Z$ are shown along a rectangular path through the phase diagram in \pref{fig:multiplets_noBM_fullBW}.

The first segment of this path corresponds to constant $U=1.5$\,eV and increasing $J$ from 0.5\,eV to 2.0\,eV, i.e., starting from the LS metallic, then to the HS metallic, and then into the CDI phase.
In the LS metallic phase for $J=0.5$\,eV, all multiplets with $N=3$, $4$, and $5$ are visited, but the probabilities of the HS states (maximal $|m_s|$) for $N=4$ and 5 are close to zero. In the transition region between LS and HS metallic phases at $J=0.75$\,eV, the HS multiplets with $|m_\mathrm{s}| = N/2$ for $N=4$ and $N=5$ become more populated, while for $J=1$\,eV, the HS states amount to nearly the total probabilities.

For increasing $J$, the probability for $N=4$ decreases linearly while the probabilities for $N=3$ and $N=5$ increase accordingly. When the system enters the CDI phase ($J>1.5$\,eV), the probabilities on the two inequivalent sites become different, but the averages over both sites, shown in \pref{fig:multiplets_noBM_fullBW}, continue their linear trend. 
In particular, even deep inside the CDI phase, for $J=2$\,eV, the probability of $N=4$ remains relatively large on both sites, indicating significant fluctuations between multiplets with different $N$, even in the insulating state.
We note that the presence of such fluctuations and the finite probability of the $N=4$ multiplet is consistent with the gradual increase of the occupation difference $\delta n$ within the CDI phase shown in \pref{fig:characteristics_noBM_fullBW}.

\begin{figure}
    \centering
    \includegraphics[width=1\linewidth]{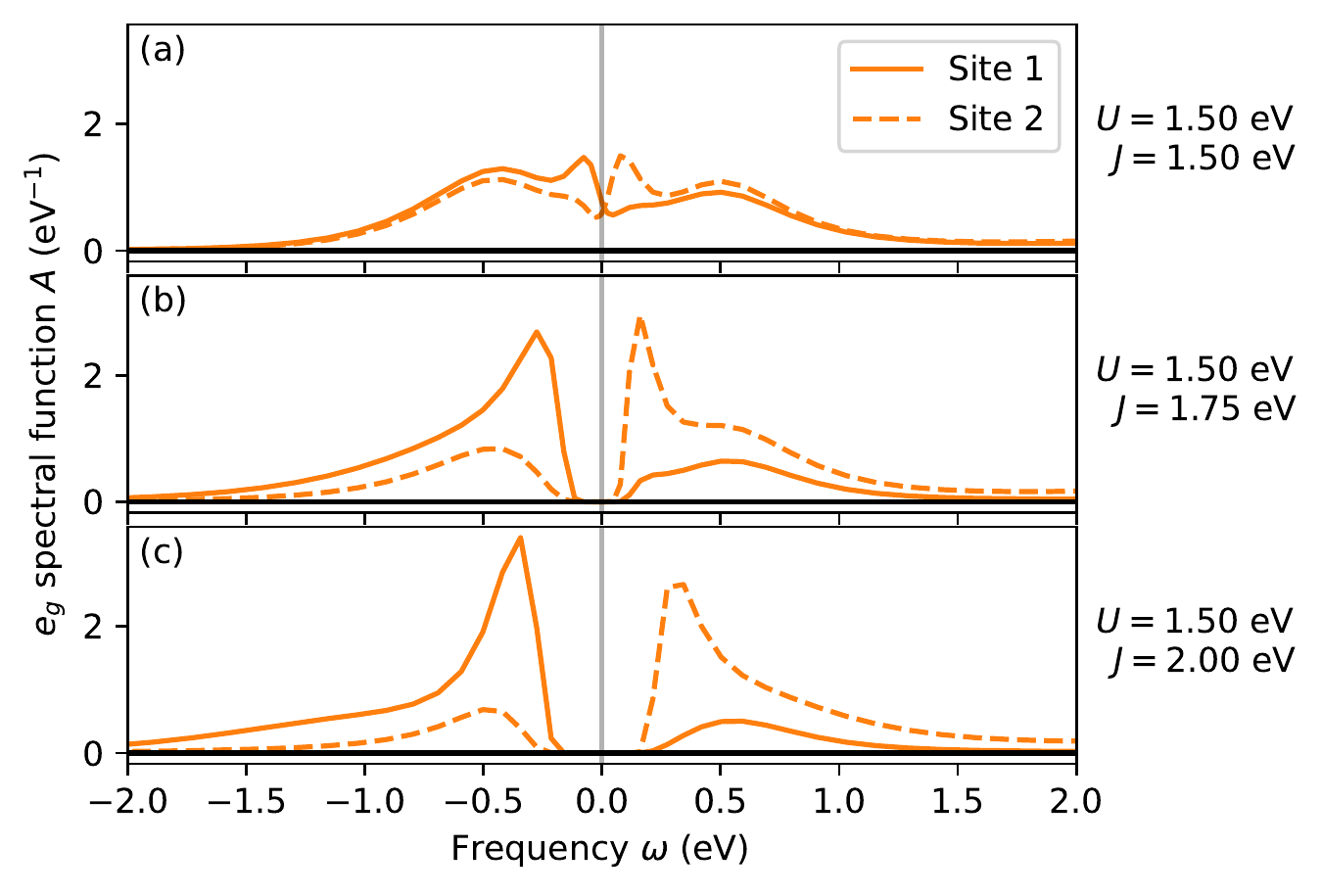}
    \caption{Site-resolved \eg spectral functions for the transition from HS metallic (top) to CDI phase (middle and bottom). A gap opens for $J \geq 1.75$\,eV. Site 1 is the more occupied one. However, both sites exhibit spectral weight above as well as below the gap (albeit with clearly different strength).}
    \label{fig:spectral_transition_CDI}
\end{figure}

\pref{fig:spectral_transition_CDI} shows the site-resolved \eg spectral functions in this regime ($U=1.5$\,eV and $J \geq 1.5$\,eV).
The spectral functions for $J>1.5$\,eV exhibit a clear gap, consistent with the zero spectral weight in \pref{fig:characteristics_noBM_fullBW}, despite the system not being in a pure local charge state, as follows from the multiplet probabilities. This differs from a recent study using a slave-boson method~\cite{isidori_charge_2019}, where the Hund's insulator in the two-orbital model was characterized by a complete charge disproportionation where only the nominal charge multiplets are populated. It also indicates that the transition to the CDI state cannot be understood purely in terms of atomic-limit considerations such as the ones in \pref{sec:atomic-limit}.

Increasing $U$ from 1.5\,eV to 6.0\,eV for fixed $J=1.5$\,eV (second segment in \pref{fig:multiplets_noBM_fullBW}) results in a transition from the CDI phase back to the HS metallic region and then moves the system towards the Mott insulating region. Here, the $N=4$ probability increases again, while the $N=3$ and $N=5$ probabilities decrease accordingly, and become zero at the boundary to the Mott insulator. Also, the quasiparticle weight $Z$ decreases continuously towards zero when approaching the Mott phase, while it remains quite large even very close to the CDI phase boundary. However, as discussed above, it is unclear whether $[1-\partial \Im \Sigma/\partial(\iu\omega)]^{-1}$ is really a good quantitative measure for the quasiparticle renormalization, here. Nevertheless, we note that a rather sharp drop of $Z$ on approaching the Hund's insulator (equivalent to our CDI phase) has also been observed in Ref.~\onlinecite{isidori_charge_2019} for a three-orbital model. 

Finally, when crossing the HS/LS transition within the Mott insulating region, the HS $N=4$ multiplet vanishes abruptly, and then, when leaving the Mott phase towards the LS metallic region, the $N=3$ and $4$ multiplets appear again, together with a gradual increase of $Z$.

Thus, to preliminary summarize the results in this section, we emphasize that we establish the presence of a spontaneously charge-disproportionated insulating phase, for strong $J$ and intermediate values of $U$, in the cubic five-orbital multi-band Hubbard model with an average of four electrons per site, analogous to results obtained for similar two- and three-orbital models~\cite{subedi_low-energy_2015, isidori_charge_2019}.
The CDI phase is characterized by a gap opening at zero frequency, but still exhibits a strong local mixing of multiplets with different number of electrons $N$. This is in contrast to the Mott insulating phase, which can be characterized by a unique local multiplet with $N=4$.
In addition to the metallic and two distinct insulating phases, a LS/HS transition occurs. This transition is very sharp inside the Mott-insulating regime but a small intermediate spin region can be observed for small $U$ in the metallic phase.

\subsection{Introducing an intrinsic site splitting and effect of reduced band width}

\begin{figure}
    \centering
    \includegraphics[width=1\linewidth]{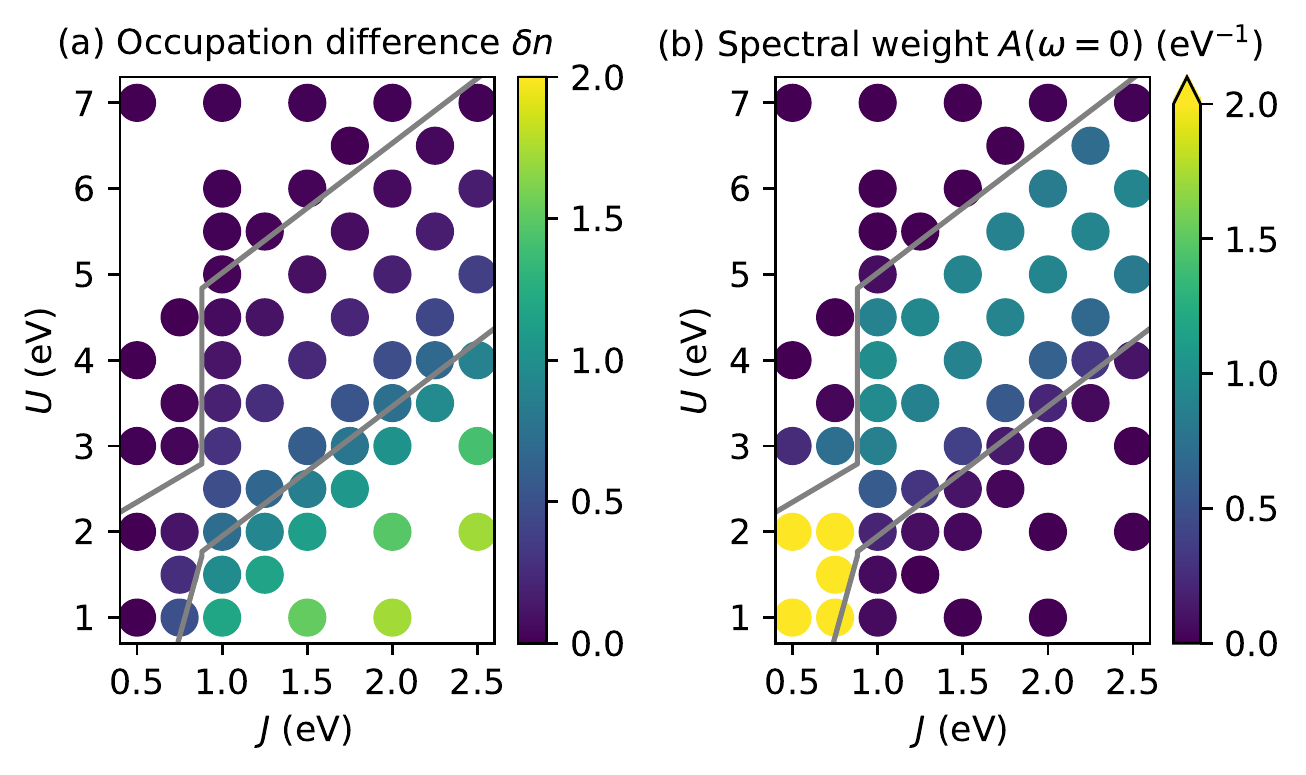}
    \caption{(a) Occupation difference $\delta n$ between inequivalent sublattices and (b) spectral weight $A(0)$ for the case with an intrinsic site splitting $\Delta_{e_g}$, plotted as function of the interaction parameters $U$ and $J$.}
    \label{fig:characteristics_BM_fullBW}
\end{figure}

Next, we investigate the effect of an intrinsic energy difference between the \eg levels on the two sublattices, which is introduced, for example, by the structural distortion that generally accompanies the charge disproportionation, i.e., the \BM breathing mode of the oxygen octahedra surrounding the transition-metal sites in the perovskite structure. We use a splitting of $\Delta_\eg = 0.436$\,eV as obtained from our fit to the band structure of CaFeO$_3$ described in \pref{sec:fitting}.

\pref{fig:characteristics_BM_fullBW} shows the spectral weight $A(0)$ and occupation difference $\delta n$ for this case. It can be seen that the region corresponding to the CDI phase is significantly increased compared to the case with $\Delta_\eg=0$ in \pref{fig:characteristics_noBM_fullBW}. The transition to the insulating state now occurs already very close to the homogeneous/inhomogeneous transition line estimated from the atomic limit. Similar behavior has also been observed for the two-band model~\cite{subedi_low-energy_2015}.
Furthermore, due to the intrinsic energy difference, a small $\delta n > 0$ already develops in the metallic phase. 
On the other hand, this small occupation difference is completely suppressed when approaching the Mott-insulating phase, and the corresponding MIT occurs again very close to the atomic limit predictions based on the $e_g$ band width. Note that the corresponding lines in \pref{fig:characteristics_BM_fullBW} are shifted relative to those in \pref{fig:characteristics_noBM_fullBW} for $\Delta_\eg=0$ as indicated in the schematic phase diagram, \pref{fig:schematic_phase_diagram}.

Due to the large extent of the CDI phase, there is now a direct competition between the CDI and the homogeneous LS metallic phase for small $J$. Nevertheless, also in this case, the corresponding phase boundary appears to be rather well described by the atomic limit prediction (gray lines in \pref{fig:characteristics_BM_fullBW}). Similar to the case with $\Delta_{e_g}=0$, we obtain a small intermediate-spin region (without significant charge disproportionation) for $U\leq 2$\,eV and $J=0.75$\,eV (data not shown in \pref{fig:characteristics_BM_fullBW}).

\begin{figure}
    \centering
    \includegraphics[width=1\linewidth]{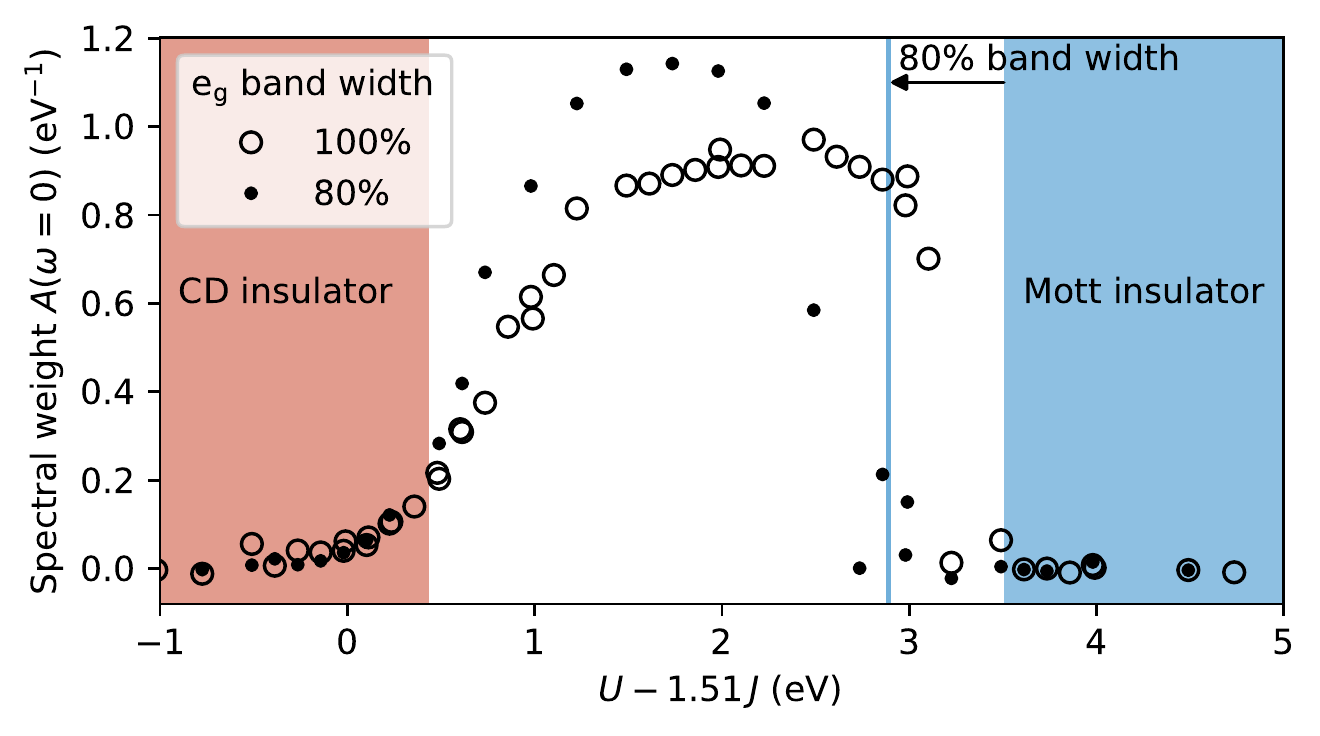}
    \caption{Effect of an 80\,\% reduction of the $e_g$ band width on the spectral weight $A(0)$ for the case with nonzero intrinsic site splitting. Only data for the HS region with $n_\eg > 0.5$ is shown as function of the ``effective'' interaction $U-1.51J$. The colored background and the vertical blue line indicate the atomic-limit predictions for the two distinct insulating phases, analogous to Fig.~\ref{fig:schematic_phase_diagram}, with the white metallic region in between.}
    \label{fig:comp_spectralw_fullredbw}
\end{figure}

Finally, we also assess the effect of reducing the $e_g$ band width, while keeping the local crystal field and the inter-site \eg splitting constant. 
To simplify the corresponding analysis, we use the observation that within the HS region ($J \geq 1$\,eV), the gradient of all quantities depicted in the $U$-$J$ phase diagrams in \pref{fig:characteristics_noBM_fullBW} and \pref{fig:characteristics_BM_fullBW} appears to be perpendicular to the two transition lines bordering the metallic phase, i.e., perpendicular to lines with $U - 1.51 J = \text{constant}$. In \pref{fig:comp_spectralw_fullredbw}, we therefore plot all available data points for the spectral weight $A(0)$ within the HS region of the phase diagram, both for full and reduced band width, as a function of the ``effective'' interaction parameter $U - 1.51 J$ (which corresponds to the difference in the local interaction energies between $d^4 + d^4$ and $d^3 + d^5$).

Indeed, all data points plotted in this way essentially condense to a single line.
In agreement with the simple atomic-limit prediction (indicated in \pref{fig:comp_spectralw_fullredbw} as vertical line or shaded regions), the transition to the Mott-insulating phase is shifted towards smaller interaction parameters if the band width is reduced, while the transition to the CD phase is barely affected by this change.
However, the occupation difference $\delta n$ in the CD region (not shown) is slightly increased for the case with reduced band width. 
Furthermore, one can also see that spectral weight disappears only very gradually 
towards the CDI state, while it disappears rather abruptly at the Mott transition.

We note that, even though the reduction of the $e_g$ band width does not significantly affect the CDI phase boundary in our calculations,  the band width could nevertheless affect the coupling to the structural distortion and can thus be an important control parameter of the system. In Ref.~\onlinecite{peil_mechanism_2019}, it was shown that in the two-orbital model the band width controls the relevant electronic susceptibility and is therefore crucial for stabilizing the CD through coupling to the structural distortion. 

\subsection{Comparison with an effective two-orbital model for equivalent sites}

After establishing and analyzing the phase diagram of the full five-orbital model, we now compare this model to the simplified effective two-orbital description introduced in \pref{sec:two-orbital}. Since by construction the effective two-orbital model assumes a HS configuration on each site, we only compare the HS part of the phase diagram. Our objective is to test how qualitatively and quantitatively accurate the computationally less demanding simplified model is. 
To ensure a meaningful comparison with the full five-orbital model, the interaction parameters of the two-orbital model are always chosen to be consistent with the corresponding parameters of the full model, as described in \pref{sec:two-orbital}. Furthermore, we only present the case with $\Delta_{e_g}=0$, i.e., without an intrinsic $e_g$ site splitting between the two sublattices. Our test calculations (not shown) indicate that a site splitting does not alter the quality of the comparison between the full five-orbital model and the effective two-orbital case.

\begin{figure}
    \centering
    \includegraphics[width=1\linewidth]{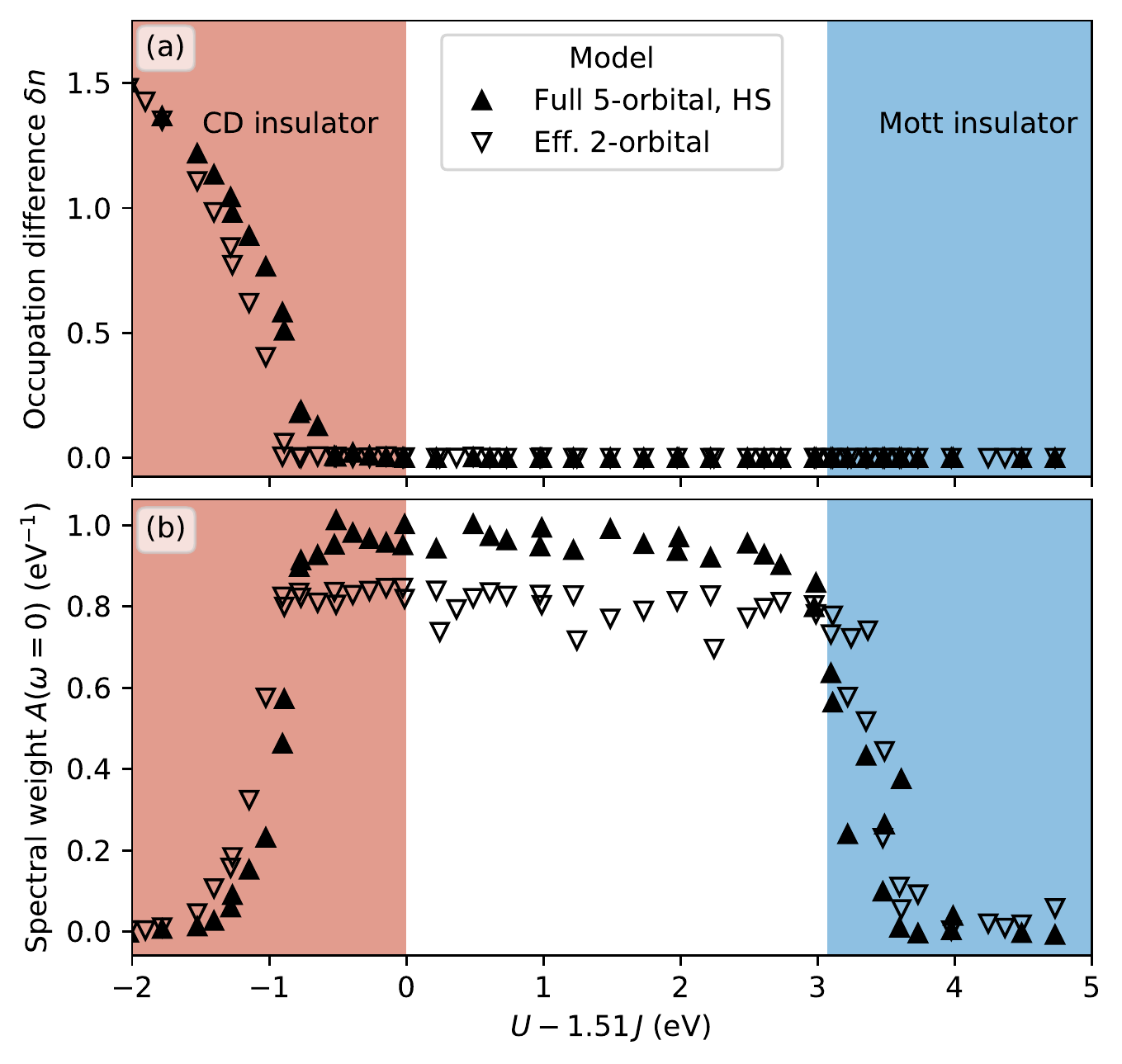}
    \caption{Comparison between the simplified effective two-orbital model and the full five-orbital description in the HS region for the case without an intrinsic site splitting, i.e., $\Delta_{e_g}=0$. (a) The occupation difference $\delta n$ between inequivalent sublattices and (b) the spectral weight $A(0)$ are shown as function of the effective interaction parameter $U - 1.51 J$. Shaded regions indicate the CD and Mott insulating regions predicted from the atomic limit (\pref{sec:atomic-limit}), white background indicates the metallic region.}
    \label{fig:characteristics_eff2_noBM_fullBW_version2}
\end{figure}

\pref{fig:characteristics_eff2_noBM_fullBW_version2} depicts the occupation difference $\delta n$ and spectral weight $A(0)$ as function of the effective interaction $U-1.51J$, both for the effective two-orbital model and the full five-orbital model. Similar to the case presented in \pref{fig:comp_spectralw_fullredbw}, all points for the same data set, obtained for different combinations of $U$ and $J$, essentially reduce to a single line, demonstrating that $U-1.51 J$ is indeed the relevant interaction parameter characterizing the behavior of the model in the HS regime.

Both models give very similar results, with only two small quantitative differences. First, the transition to the CDI phase occurs for slightly smaller effective interaction in the simplified two-orbital model. Due to the large increase of $\delta n$ at the transition, this can lead to quantitative differences in the charge disproportionation of up to 0.5 electrons, but only in a very small region close to the phase transition. Second, the spectral weight at zero energy in the metallic phase is slightly reduced in the effective two-orbital model. 

\begin{figure*}
    \centering
    \includegraphics[width=1\textwidth]{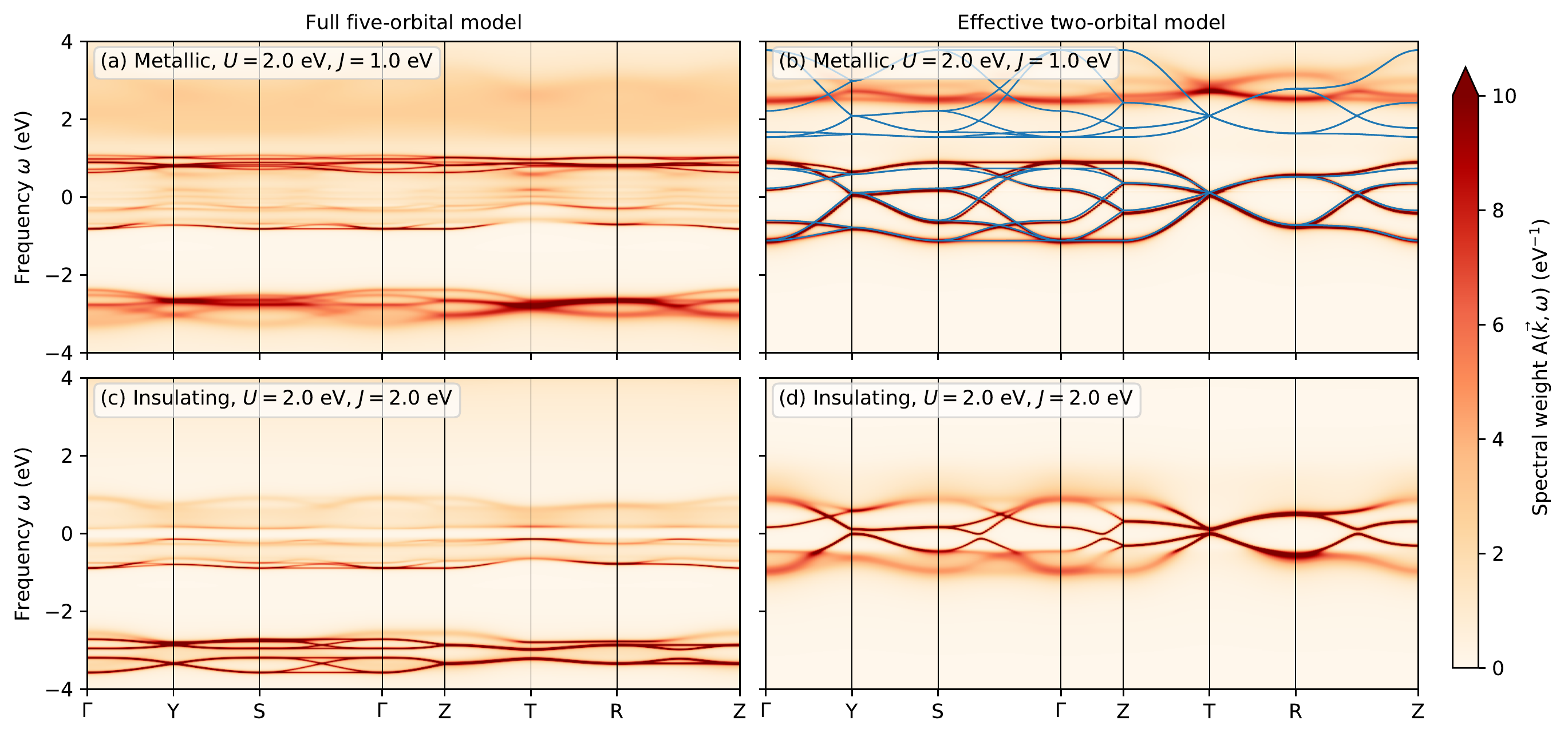}
    \caption{Comparison of $k$-resolved spectral functions between the full five-orbital model (a and c) and the simplified effective two-orbital model (b and d) in the HS metallic state close to the HS/LS transition (a and b) and in the CD insulating state (c and d). The blue lines in (b) are the bands obtained from the non-interacting part of the effective model, after taking into account the local spin splitting as well as the renormalized and spin-mixed hopping matrices.}
    \label{fig:spagh_comp}
\end{figure*}

Thus, it appears that the basic observables, such as occupations and the resulting phase boundaries, as well as the gap opening at the Mott transition, obtained from the simplified two-orbital model agree well with the corresponding results for the full five-orbital model. However, this is not true for more complex spectral properties, as can be seen from the $k$-resolved spectral functions shown in \pref{fig:spagh_comp} (obtained through analytic continuation of the self-energy).

For the full five-orbital model, the occupied and unoccupied $t_{2g}$ bands appear as rather sharp features. For the case with $U=2.0$\,eV and $J=1.0$\,eV these bands are situated at around $-$3\,eV and 1\,eV.  
The $e_g$ spectral weight is much more incoherent, with significant lifetime broadening due to many-body correlations. In the metallic state obtained for $J=1.0$\,eV, some dispersion is still visible between $\omega = \SI{-1}{eV}$ and $\SI{1}{eV}$. At higher energies, a broad incoherent spectral background is present at around 2\,eV.

In contrast, the $k$-resolved spectral functions obtained from the effective two-orbital model exhibit much weaker many-body effects. In the metallic regime for $J=1.0$\,eV, a sharp band structure is visible, in particular around $\omega=0$. The band dispersion is essentially identical to the one following from the renormalized and spin-mixed hopping amplitudes, also shown as blue lines for comparison in \pref{fig:spagh_comp}(b). The Coulomb interaction between the $e_g$ electrons only leads to a weak lifetime broadening of the individual bands and mostly increases the splitting between local majority and minority spin channels. 
Close to the CDI transition  at $J=2$\,eV, the local minority spin bands are shifted to even higher energies and a gap starts opening at zero frequency, consistent with the full five-orbital model. However, in the two-orbital model, the bands around the gap seem to remain rather well defined, while the band edges at approximately $\pm 1$\,eV become rather fuzzy, which does not happen in the five-orbital case.

The fact that the effective two-orbital model exhibits much weaker many-body correlations than the full five-orbital model is to be expected, since the interaction with the $t_{2g}$ core spins merely enters via the renormalized hopping amplitudes and in the form of ``classical'' Zeeman fields. However, the local self-energies also seem to exhibit some non-trivial differences, leading to pronounced lifetime broadening in different energy regions. Nevertheless, these differences do not seem to critically affect integrated quantities such as occupations or the spectral weight around zero energy (see \pref{fig:characteristics_eff2_noBM_fullBW_version2}).

\section{Conclusions and Summary}

In summary, we have established the existence of a charge-disproportionated insulating phase (or Hund's insulator) within a generic cubic five-orbital full $d$-shell model and an average filling of four electrons per site. Our results are consistent with previous studies of two- and three-orbital models~\cite{subedi_low-energy_2015, isidori_charge_2019}, which all show the same overall phase diagram with two distinct insulating phases, the standard homogeneous Mott insulator for large $U$ and a spontaneously charge disproportionated insulator for large $J$ and moderate $U$, separated by a metallic region.

For the present case of the cubic five-orbital model, additional complexity arises from the presence of the LS/HS transition. In particular for $U<2$\,eV and $J\lesssim 1$\,eV, the metallic LS state competes with the CDI state. We note that this parameter regime is realistic for materials such as ferrates, where the Coulomb repulsion between the $d$ electrons can be strongly screened due to the (potentially negative) charge-transfer character and the resulting strong hybridization with the oxygen $p$ bands. 
Furthermore, it can be expected that the incorporation of intersite Coulomb interactions, neglected in the present study to focus on the effect of the local Hund's coupling, will also favor the charge disproportionated state and thus further reduce the minimal $J$ required to obtain the CDI phase for a given $U$~\cite{seth_renormalization_2017}.

Our results also show that the CDI phase is strongly affected by a small energy splitting of the $e_g$ levels between the charge-disproportionated sublattices, introduced here to mimic the effect of an additional structural distortion that accompanies the CD. Again, this is analogous to previous results obtained for the two-orbital case~\cite{subedi_low-energy_2015}, and shows that the CDI couples very strongly to such a structural distortion. Thus, for quantitative studies of specific materials, it is crucial to account for this coupling and treat the structural distortion as a free variable~\cite{peil_mechanism_2019}.

An interesting aspect is also the unconventional nature of the CDI state. While for the case of the homogeneous Mott insulator, the critical $U$ and $J$ values for the MIT can be estimated with rather good accuracy from the atomic limit, this is not the case for the transition to the (spontaneous) CDI state. 
Our results show that a gap in the spectral function opens even though there are still significant fluctuations between different local multiplets. This is also consistent with the gradual increase of the charge disproportionation. Nevertheless, for the case with a nonzero site splitting $\Delta_{e_g}$, the homogeneous/CD phase boundary is well approximated by the corresponding atomic limit. 

Finally, we have also compared the full five-orbital model with a simplified effective two-orbital description applicable to the HS limit. The simplified model gives results very similar to the full model in the HS region, even though the transition to the CDI state occurs at slightly lower effective interaction $U - 1.51 J$ than in the full model. In spite of clear differences in the $k$-resolved spectral functions, integrated quantities such as occupations or the zero-energy spectral weight $A(\omega=0)$ agree very well between the two models.
However, assuming a realistic parameter regime for perovskite transition-metal oxides such as CaFeO$_3$ around $J\lesssim 1$\,eV and $U=$\ 1-2\,eV, our calculated phase diagrams indicates that these materials could be rather close to the LS/HS transition, and therefore might require a treatment within the full model.

Our work shows that the formation of a charge-disproportionated insulator, in a regime where the Hund's coupling outweighs the strongly screened Coulomb repulsion, is a general feature of multi-band Hubbard models at specific filling levels. Based on this observation, our rather generic cubic five-orbital model represents a good starting point and can serve as reference for future studies of specific materials including, e.g., ferrates and manganites.

\appendix*
\begin{acknowledgments}
We thank Sophie Beck and Alexander Hampel for their help regarding the use of soliDMFT and TRIQS/DFTTools as well as for many enlightening discussions, and Alexander Gillmann, who performed initial calculations related to this project for his BSc thesis.
This work was supported by ETH Zurich and the Swiss National Science Foundation through NCCR-MARVEL. Calculations have been performed on the cluster ``Piz Daint'' hosted by the Swiss National Supercomputing Centre and supported under project IDs s889 (User Lab) and
mr26 (MARVEL).
\end{acknowledgments}

\appendix
\section{Atomic-limit calculations} \label{app:atomic_limit}

The local interaction energies for all low-energy configurations (both LS and HS) with $2 \leq N \leq 6$ electrons per site, corresponding to the interaction Hamiltonian in \pref{eq:hint} are listed in \pref{tab:energies}. The corresponding energies can be expressed exactly in terms of the Slater integrals $F_0$, $F_2$, and $F_4$ (see, e.g., Ref.~\onlinecite{pavarini_ldadmft_2011}), while for the expression in terms of $U$ and $J$ we use the definitions described in \pref{sec:ham}.
The full local energies $E_N^{\text{(LS/HS)}}$ in the atomic limit are then easily obtained by adding multiples of the crystal-field energy $\Delta_{\eg-\ttg} \pm \Delta_{\eg}/2$ according to the corresponding $e_g$ occupation and the relevant sublattice. 

\begin{table}
\centering
\caption{Interaction energies of the lowest HS/LS $N$ electron configuration (for $2\leq N \leq 6$), corresponding to the Hamiltonian in \pref{eq:hint}, expressed both in terms of the Slater integrals $F_0$, $F_2$, $F_4$, and in terms of the interaction parameters $U$ and $J$. Note that for $N=2$ and $N=3$ there is no ambiguity between HS and LS. }
\label{tab:energies}
\begin{ruledtabular}
\begin{tabular}{lcc}
$N=2$ & $F_0 - 5/49 F_2 - 24/441 F_4$ & $U - 1.171 J$ \\
$N=3$ & $3 F_0 - 15/49 F_2 - 72/441 F_4$ & $3 U - 3.513 J$\\
$N=4$, LS & $6 F_0 - 15/49 F_2 - 44/441 F_4$ & $6 U - 3.169 J$ \\
$N=4$, HS & $6 F_0 - 21/49 F_2 - 189/441 F_4$ & $6 U - 6 J$ \\
$N=5$, LS & $10 F_0 - 20/49 F_2 - 40/441 F_4$ & $10 U - 3.997 J$ \\
$N=5$, HS & $10 F_0 - 35/49 F_2 - 315/441 F_4$ & $10 U - 10 J$ \\
$N=6$, LS & $15 F_0 - 30/49 F_2 - 60/441 F_4$ & $15 U - 5.995 J$ \\
$N=6$, HS & $15 F_0 - 35/49 F_2 - 315/441 F_4$ & $15 U - 10 J$
\end{tabular}
\end{ruledtabular}
\end{table}

The transition boundary between the homogeneous $d^4$ HS and LS states is then obtained by comparing the corresponding energies:
\begin{align}
    0 &= E_4^\mathrm{(LS)} - E_4^\mathrm{(HS)} \nonumber\\
    &= \frac6{49} F_2 + \frac{145}{441} F_4 - \Delta_{\eg - \ttg} \nonumber\\
    &\approx 2.83 J - \Delta_{\eg - \ttg}
\end{align}

For the transition between the inhomogeneous CD region and the homogeneous HS and LS regions, respectively, we compare the corresponding energies per two sites:
\begin{align}
    0 &= \left(E_5^\mathrm{(HS)} + E_3\right) - 2 E_4^\mathrm{(HS)} \nonumber\\
      &= F_0 - \frac8{49} F_2 - \frac1{49} F_4 - \Delta_\eg          \nonumber \\
      &\approx U - 1.51 J - \Delta_\eg \quad ,
      \label{eq:phase-boundary1}
\end{align}
and:
\begin{align}
    0 &= \left(E_5^\mathrm{(HS)} + E_3\right) - 2 E_4^\mathrm{(LS)} \nonumber\\
      &= F_0 - \frac{20}{49} F_2 - \frac{299}{441} F_4 + 2(\Delta_{\eg - \ttg} - \Delta_\eg / 2) \nonumber\\
      &\approx U - 7.17 J + 2\Delta_{\eg - \ttg} - \Delta_\eg \quad .
\end{align}
Here, of course, we assume that in the presence of an \eg site splitting, the HS $d^5$ state is formed on the site with the lower \eg energy.

The border of the Mott insulating regions within the homogeneous HS and LS phases are estimated by comparing the energy of the lowest inter-site charge transfer excitation ($2d^4 \rightarrow d^5+d^3$) with the relevant band width. Note that for the homogeneous HS region, this results in the same energy terms as in \pref{eq:phase-boundary1}, $E_5^\mathrm{(HS)} + E_3 - 2 E_4^\mathrm{(HS)}$, but now this is set equal to the \eg band width $W_\eg$.
The same argument is also applied to identify the metal-insulator boundary within the homogeneous LS phase, where the electrons are restricted to states within the \ttg sub-manifold and therefore the inter-site charge transfer energy is compared to the \ttg band width $W_\ttg$:
\begin{align}
\label{eq:Mott-LS}
    W_\ttg &= \left(E_5^\mathrm{(LS)} + E_3\right) - 2 E_4^\mathrm{(LS)} \nonumber\\
    &= F_0 - \frac5{49} F_2 - \frac{24}{441} F_4 \nonumber\\ 
    &\approx U - 1.17 J \quad .
\end{align}
Note that if the right side of \pref{eq:Mott-LS} is set to zero, this also defines a potential transition to an inhomogeneous LS phase, which is indeed more favorable than the HS CD state for $J< \Delta_{\eg- \ttg}/3$. However, with $\Delta_{\eg - \ttg} \approx 2.5$\,eV, this requires $J \lesssim 0.83$\,eV and simultaneously $U \lesssim 1.17J \lesssim 0.98$\,eV. Since we do not consider this parameter range in our DMFT calculations, we neglect the corresponding phase also in the schematic phase diagram  in \pref{fig:schematic_phase_diagram}.

The estimation of the metal-insulator boundary within the inhomogeneous CD region is less straightforward. As mentioned in \pref{sec:atomic-limit}, the effective kinetic energy required to destabilize the CDI state towards the homogeneous metal cannot be estimated from simple band-width considerations (at least for $\Delta_{e_g}=0$).   
In addition, the CDI state can be destabilized towards low $U$ values through an effective kinetic hopping within only one specific sublattice. As pointed out in Ref.~\onlinecite{subedi_low-energy_2015}, this involves indirect hopping processes, which again complicates a simple band-width estimation of the corresponding kinetic energy. 

For the case with hopping only among the nominal $d^5$ sites ($2d^5 \rightarrow d^4 + d^6$), one obtains:
\begin{align}
\label{eq:previous}
    (W_\ttg + W_\eg)/2 &\leq \left(E_6^\mathrm{(HS)} + E_4^\mathrm{(HS)}\right) - 2 E_5^\mathrm{(HS)} \nonumber\\
    &= F_0 + \frac{14}{49} F_2 + \frac{126}{441} F_4 \nonumber\\&\quad- (\Delta_{\eg - \ttg} - \Delta_\eg/2) \nonumber\\
    &= U + 4 J - \Delta_{\eg - \ttg} + \Delta_\eg/2
\end{align}
Note that here we have used the average band width as an upper limit for the effective kinetic energy, which corresponds to the red-hatched area around $J\approx 1$\,eV in \pref{fig:schematic_phase_diagram}. The true effective band width corresponds to an indirect hopping between $d^5$ sites and is, therefore, expected to be significantly smaller, and thus this transition barely plays a role in our considerations. We can also identify $(W_\ttg + W_\eg)/2 + \Delta_{\eg - \ttg} \equiv W_\mathrm{full-d}$ leading to
\begin{align}
    W_\mathrm{full-d} - \Delta_\eg/2 \leq U + 4 J
\end{align}

For the case with hopping only between nominal $d^3$ sites, and assuming that the involved $d^4$ state is LS, one obtains (again using the full \ttg band width as upper limit):
\begin{align}
    W_\ttg &\leq \left(E_4^\mathrm{(LS)} + E_2\right) - 2 E_3 \nonumber\\
    &= F_0 + \frac{10}{49} F_2 + \frac{76}{441} F_4 \nonumber \\
    &\approx U + 2.69 J
\end{align}
This only predicts a metallic state in the region where the hopping between $d^5$ sites is already active, and is therefore not relevant. However, if one considers also hopping through a $d^4$ HS state, and using an average band width as in \pref{eq:previous}, one obtains:
\begin{align}
    (W_\ttg + W_\eg)/2 &\leq \left(E_4^\mathrm{(HS)} + E_2\right) - 2 E_3 \nonumber\\
    &= F_0 + \frac4{49} F_2 - \frac{69}{441} F_4 \nonumber\\ &\quad+ (\Delta_{\eg - \ttg} + \Delta_\eg/2) \nonumber\\
    &\approx U - 0.15 J + \Delta_{\eg - \ttg} + \Delta_\eg/2
\end{align}
This can in principle destabilize the CDI state for very low $U$ and very high $J$ and is also indicated by the red-hatched area in \pref{fig:schematic_phase_diagram}. 

Finally, we note that none of the multiplet energies listed above would be affected by the spin-flip and pair-hopping terms that we ignore within the density-density approximation, \pref{eq:hint}. These additional terms would only affect higher energy multiplets, which are not crucial for a correct description of the ground state phase diagram, as they do not affect the phase boundaries obtained from this atomic limit consideration. This provides a good {\it a priori} justification for our use of the density-density approximation.

\bibliography{bibfile}

\end{document}